\documentstyle[amsmath,amssymb,graphicx]{article}

\def\be{\begin{eqnarray}}
\def\ee{\end{eqnarray}}
\def\nn{\nonumber}

\def\p{\partial}
\def\tr{{\rm tr}\,}

\def\[{\phantom.[}
\def\lb{\left[}
\def\rb{\right]}

\hoffset= - 1.0in         

\title{\Large {\bf Integrability properties of Hurwitz partition
functions. \\ II. Multiplication 
of cut-and-join operators and WDVV equations}
\vspace{.2cm}}
\author{{\bf A.Mironov}\footnote{ {\small {\it
Lebedev Physics Institute} and {\it ITEP, Moscow, Russia}};
mironov@itep.ru; mironov@lpi.ru}, {\bf A.Morozov}\thanks{{\small
{\it ITEP, Moscow, Russia}}; morozov@itep.ru} \ and {\bf
S.Natanzon}\thanks{{\small {\it
Department of Mathematics, Higher School of Economics, Moscow, Russia};
{\it ITEP, Moscow, Russia}  and
{\it A.N.Belozersky Institute, Moscow State University, Moscow, Russia};
natanzons@mail.ru}}\date{ }}

\begin{document}

\maketitle

\vspace{-6.0cm}

\begin{center}
\hfill FIAN/TD-14/11\\
\hfill ITEP/TH-24/11\\
\end{center}

\vspace{4.5cm}

\begin{abstract}
Correlators in topological theories are given by the values of a
linear form on the products of operators from a commutative
associative algebra (CAA). As a corollary, partition functions of
topological theory always satisfy the generalized WDVV equations of
\cite{MMM}. We consider the Hurwitz partition functions,
associated in this way with the CAA of cut-and-join operators. The
ordinary Hurwitz numbers for a given number of sheets in the
covering provide trivial (sums of exponentials) solutions to the
WDVV equations, with finite number of time-variables. The
generalized Hurwitz numbers from \cite{MMN} provide a non-trivial
solution with infinite number of times. The simplest solution of
this type is associated with a subring, generated by the
dilatation operators $\hat W_1 = \tr D = \tr X\p/\p X$.
\end{abstract}

\section{Introduction}

Hurwitz theory is one of the rapidly developing branches
of modern mathematical physics \cite{Hurfirst}-\cite{I}.
It has its origins in the enumeration problem
of ramified coverings of $CP^1$,
and it is brought into modern context
with the formula due to Frobenius, \cite{Dijk}
\be
{\rm Cover}_n(\Delta_1,\ldots,\Delta_m)
= \sum_{R} d^2_R \varphi_R(\Delta_1)\ldots\varphi_R(\Delta_m)
\delta_{|R|,n}
\label{Frof}
\ee
expressing  the covering multiplicities through the quantities
$\varphi_R(\Delta)$ proportional to the characters of the symmetric group $S_n$
\cite{Mac}.
Here $\Delta_i$ is the Young diagram (integer partition),
characterizing the type (conjugation class) of ramification point,
all sizes of the diagrams being the same and equal to
the number of sheets in the covering, $|\Delta_1| = \ldots =
|\Delta_m|=n$,
and the sum runs over the Young diagrams $R$ of the same size
$|R|=n$, but this time they label representations of the symmetric
group.
The symmetric group characters are among the most important objects
in combinatorics, more sophisticated then the
$GL(\infty)$ characters $\chi_R(t)$, very well-known and used
in physical applications \cite{Ham}.
Still the two sets of characters are directly related by the Frobenius formula,
which is a sort of Fourier transform \cite{Mac},
\be
\chi_R(t) = \sum_\Delta d_R\varphi_R(\Delta) p(\Delta)
\delta_{|\Delta|,|R|}
\label{chiphitr}
\ee
where $\Delta = [\ldots\geq\delta_2\geq\delta_1] =
[\ldots , 3^{m_3},2^{m_2},1^{m_1}]$, its size
$|\Delta| = \sum_k \delta_k = \sum_k km_k$,
the time monomial
\be
p(\Delta) = \prod_k p_{\delta_k} = \prod_k p_k^{m_k}
\ee
and $p_k = kt_k$.
These two character sets are further unified through the
concept \cite{GJV} of cut-and-join operators $\hat W(\Delta)$,
commuting
differential operators of degree $|\Delta|$ in $t$-variables,
for which they serve as eigenvectors and eigenvalues respectively
\cite{MMN}:
\be
\hat W(\Delta)\chi_R(t) = \varphi_R(\Delta)\chi_R(t)
\label{WchiI}
\ee
The problem with these operators is that they look rather
complicated (they belong to the class of $W$-operators,
made from powers of the $U(1)$ current).
In \cite{MMN} we explained that the cut-and-join operators acquire
a very simple form if expressed in terms of the matrix Miwa variables
$p_k = \tr\ X^k$, then\footnote{
The combinatorial coefficient $z(\Delta)=
\prod_k \frac{1}{k^{m_k}m_k!}$ here that counts the order of the
automorphism group of the Young diagram, appears everywhere in
the  theory of symmetric functions and symmetric group $S(\infty)$.
In particular, the standardly normalized symmetric group characters
$\hat\chi_R(\Delta) = d_R z(\Delta)\varphi_R(\Delta)$.
They are generated by the command
${\large Chi(R, \Delta)}$ in Maple in the package
{\large combinat}.
}
\be\label{caj}
\hat W_\Delta = \ : \prod_k \frac{1}{k^{m_k}m_k!}\big(\tr \hat D^k)^{m_k}\, :
\ee
for the $GL(\infty)$ matrix generator $D = X\frac{\p}{\p X_{{\rm tr}}}$.
This representation opens a constructive way to evaluation
of the structure constants $C_{\Delta_1\Delta_2}^\Delta$,
which appear in the CAA of cut-and-join operators and,
as a consequence, in the ordinary multiplication algebra
of symmetric group characters (this algebra was earlier considered from
combinatorial point of view in \cite{IK} where it is claimed to be
equivalent to the algebra of shifted symmetric functions of \cite{OO}):
\be
\hat W_{\Delta_1} \hat W_{\Delta_2} =
\sum_\Delta C_{\Delta_1\Delta_2}^\Delta \hat W_\Delta, \nn \\
\varphi_R(\Delta_1)\varphi_R(\Delta_2)
= \sum_\Delta C_{\Delta_1\Delta_2}^\Delta \varphi_R(\Delta)
\label{CAAW}
\ee
It is important in these formulae that the sums over $\Delta$
are not restricted to $|\Delta| = |\Delta_{1,2}|$, moreover,
$|\Delta_1|$ can be different from $|\Delta_2|$.
Note that already in (\ref{WchiI}) there is no restriction to
$|R| = |\Delta|$, and $\varphi_R(\Delta)$'s in this formula
are more general than those in (\ref{chiphitr}). They are defined for
the Young with $r$ unit rows, $[\Delta]=[\tilde\Delta,1^r]$ by adding
additional unit rows to reach $|R|=|\Delta|$ in accordance with the rule
\be
\varphi_R([\Delta]) \equiv \left\{\begin{array}{ccc}
0 & {\rm for} &  |\Delta|>|R| \\
{(|R|-|\Delta|+r)!\over r!(|R|-|\Delta|)!}\
\varphi_R([\Delta,\underbrace{1,\ldots,1}_{|R|-|\Delta|}])
& {\rm for} &  |\Delta|\leq |R|
\end{array}\right.
\label{phiext}
\ee
This naturally leads to extension of (\ref{Frof})
by removing the projector $\delta_{|R|,n}$ and lifting the restriction
$|\Delta_1| = \ldots = |\Delta_m| = n$, which defines
what was called {\it generalized} Hurwitz numbers in \cite{MMN}.

Note that the CAA of cut-and-join operators induces the multiplication on the Young diagrams:
\be
\Delta_1 * \Delta_2 =\sum_\Delta C_{\Delta_1\Delta_2}^\Delta \Delta
\ee
This multiplication can be considered as extension of another $\circ$-multiplication on the
Young diagrams, given by the composition of permutations and related to the ordinary Hurwitz
numbers. This latter is connected with the *-multiplication by
restricting onto diagrams of the same size:
\be
\Delta_1 \circ \Delta_2 =\sum_\Delta C_{\Delta_1\Delta_2}^\Delta \Delta\
\delta_{|\Delta_1|,|\Delta|}
\ee
with $|\Delta_1|=|\Delta_2|$. Inversely, one can construct $*$-multiplication from the
$\circ$-one by the procedure described in section 3.

In this paper we discuss one of the immediate implications
of the extension to the generalized Hurwitz numbers: their generating function
is the partition function of a topological field theory
associated with the CAA (\ref{CAAW}) and, hence, satisfies
the WDVV equations of \cite{MMM}.
The ordinary Hurwitz partition functions, based on the ordinary
$\circ$-multiplication in symmetric group $S(n)$, also satisfy
WDVV for each given $n$, but they provide only trivial
solutions (which, however, were not discussed in the
existing literature).
The WDVV equations are imposed on "quasiclassical $\tau$-functions",
which are obtained by one or another kind of Whitham
averaging procedure \cite{Whith} from the KP/Toda hierarchy
and a particular set of Riemann surfaces (background) with additional data.
The quasiclassical hierarchies are well studied in the case
when the background is a Riemann sphere, but in the case
of Hurwitz theory it should be different (a Lambert curve,
for example, \cite{mari,mmhk,kaz}), and such hierarchies are not yet described (see,
however, \cite{Takasaki}).
An advantage of the quasiclassical hierarchy would be that
particular equations involve only derivatives
w.r.t. the finite number of time-variables, while the WDVV equations
involve inversion of an infinite size matrix.
Derivation of such reducible equations for Hurwitz partition
functions remains an open problem.

This paper is the second in the series which describes properties of the generating
functions in Hurwitz theory. The first paper, \cite{I} contains a summary of
integrable properties, which will be described in detail in our next paper,
\cite{III}.

\bigskip

In section 2 we begin with reminding the general construction
of topological theory for any CAA and explain why its partition
function always satisfy the WDVV equations (their original
form is a little more general than in \cite{MMM}, and far
more general than in \cite{WDVV}, the triple derivative
equations of \cite{MMM} being a direct corollary, but not vice versa!).
Then, in sections 3 we provide discuss the two multiplications:
$\circ$ and *-products. The corresponding multiplication tables can be found in
Appendices I and II.
Knowledge of these tables allows one to examine concrete examples illuminating
following sections. In section 4 we construct
two types of generating functions which are associated
with two multiplications, and in section 5 the corresponding WDVV equations
satisfied by these
generating functions are discussed.
As a particular example, in sections 6 and 7
we provide details about the $[1^p]$ subring
of the *-algebra, describing cut-and-join operators,
associated with the single row diagrams
(the "complementary" single column operators would instead
generate the entire algebra).

\section{Topological theories and WDVV}

\subsection{Topological theory on sphere (tree level)}

At the tree (string) level (i.e. on sphere) a topological theory is defined by three
ingredients:

\begin{itemize}
\item a vector space with the basis of "observables"
$\{\phi_i\}$,
\item an associative and commutative multiplication
\be
\phi_i *
\phi_j = \sum_k C_{ij}^k \phi_k
\ee
\item
and a linear form ($c$-valued function) on this space
$<\phi_i> = K_i$.
\end{itemize}

At the loop (string) level (i.e. on higher genera Riemann surfaces) one needs also to define
traces and impose an additional constraint on the toric $1$-point function (in addition to
associativity and commutativity of the multiplication), but we do not need this in
the present text, which is fully devoted to the tree level topological
theories.

The tree correlators are defined as
\be
K_{i_1\ldots i_n} = <
\phi_{i_1},\ldots, \phi_{i_n}> = < \phi_{i_1}*\ldots *\phi_{i_n} > =
\sum_k C_{i_1\ldots i_n}^k <\phi_k> = \sum_k C_{i_1\ldots i_n}^k K_k
\ee
where the coefficients $C$ are products of the original 3-valent
structure constants $C_{ij}^k$. These correlators are totally
symmetric under permutations of $i_1,\ldots,i_n$.

It is also convenient to introduce "the bare metric"
\be
G_{ij} \equiv
<\phi_i,\phi_j> = <\phi_i*\phi_j> = \sum_k C_{ij}^k <\phi_k>
\ee
and use it and its inverse $G^{ij}$ to raise and lower indices, in particular, to construct
the totally symmetric tensors
\be
C_{ijk} \equiv <\phi_i,\phi_j,\phi_k> = \sum_m
C_{ij}^mG_{mk} = \sum_m C_{jk}^m G_{mi} = \sum_m C_{ik}^m G_{mj}
\ee
and
\be <\phi_i,\phi_j,\phi_k,\phi_l> = \sum_m C_{ij}^m C_{mkl} =
\sum_m C_{ik}^m C_{mjl} = \sum_m C_{il}^m C_{mjk} = \ldots
\ee
Next one defines the tree partition function
\be
Z[\beta] = \left<
e_*^{\sum_i \beta_i\phi_i} \right> = <E[\beta]> \equiv <<1>>
\ee
where $e_*(\phi) = \sum_n \frac{1}{n!}\underbrace{\phi*\phi*\ldots*\phi}_n$. Then,
\be
C_{ijk} =
\left.\frac{\p^3Z[\beta]}{\p\beta_i\p\beta_j\p\beta_k}\right|_{\beta
= 0}
\ee

\subsection{Deformation by coupling constants $\beta$
and WDVV equations}

One can now introduce deformed, $\beta$-dependent algebra with a
$\beta$-dependent multiplication
\be
\phi_i \hat * \phi_j \equiv
\phi_i *\phi_j *E[\beta] \equiv \sum_k \hat C_{ij}^k \phi_k
\ee
where $E[\beta]$ is a family of elements of the algebra. The new
multiplication
is still commutative and associative:
\be
(\phi_i\hat *\phi_j)\hat *
\phi_k = \phi_i*\phi_j*\phi_k*E[\beta]*E[\beta] = \phi_i \hat *
(\phi_j \hat * \phi_k)
\ee
It is also possible to introduce the deformed
observables $\hat\phi_i = \phi_i * E[\beta]$, then
\be
\hat\phi_i *
\hat\phi_j = \phi_i * \phi_j * E[\beta]*E[\beta] = (\phi_i \hat
*\phi_j) * E[\beta] = \widehat{(\phi_i \hat * \phi_j)} = \sum_k \hat
C_{ij}^k\hat \phi_k
\ee
is also a commutative associative algebra.

Now one can introduces the $\beta$-dependent correlators:
\be
<<
\phi_{i_1},\ldots,\phi_{i_n} >>\ \equiv\ <\phi_{i_1}*\ldots
*\phi_{i_n}* E[\beta]>
\ee
Then the triple correlator possesses {\it
two} alternative representations (the last two sums in this formula):
\be
\hat C_{ijk} \equiv \frac{\p^3
Z[\beta]}{\p\beta_i\p\beta_j\p\beta_k} =\ <<\phi_i,\phi_j,\phi_k>>\
=\ <\phi_i*\phi_j*\phi_k*E[\beta]>\ = \sum_m \hat C_{ij}^m G_{mk} =
\sum_m C_{ij}^m \hat G_{mk} \label{hatnothat}
\ee
The first representation
is in terms of deformed $\hat C_{ij}^k$ and the {\it bare} metric
$G_{mk} = <\phi_k*\phi_m>$, while the second one is in terms of the
{\it bare} (undeformed) $C_{ij}^k$ and the deformed metric
\be
\hat
G_{mk} \equiv\ <<\phi_m,\phi_k>>\ =\ <\phi_m*\phi_k*E[\beta]>\ =
\frac{\p^2 Z[\beta]}{\p\beta_k\p\beta_m}
\ee
Associativity of
original and deformed algebras is expressed in the commutativity condition
of the structure constants $\left(\check C_i\right)^k_j\equiv C_{ij}^k$
\be\label{assC}
\check C_i\check C_j=\check C_j\check C_i
\ee
In its turn, this implies that
\be
\hat C_{ijm} G^{mn}
\hat C_{kln} = \hat C_{ikm} G^{mn} \hat C_{jln}
\ee
and
\be
\hat
C_{ijm} \hat G^{mn} \hat C_{kln} = \hat C_{ikm} \hat G^{mn} \hat
C_{jln} \label{WDVVf}
\ee
which we respectively call as {\it
bare} and {\it full} \cite{MMM,gWDVV} WDVV equations for the
partition function $Z[\beta]$ (in the case of the {\it full} WDVV
equation all ingredients are triple derivatives of $Z[\beta]$,
the deformed metric is the triple correlator with the unity operator
$\phi_0 = I$: $\hat G_{mn} = \hat C_{0mn}$).

In some cases the choice of the vector space and observables
$\phi_i$ is not unique. The same algebra may possess different
realizations (representations) and one can ask if the
$\beta$-deformed topological theory respects this freedom. The
problem is very similar to the representation theory of Lie algebras and
to the concept of the universal group elements etc.

\subsection{Hurwitz topological theory}

Consider here an explicit example
when $\phi_i$ has an additional label, $R$ that is the variable averaged in the
mean value $<...>$, and such that the structure constants in the product
\be
\phi_i(R)\phi_j(R) = \sum_k C_{ij}^k \phi_k(R)
\ee
do not depend on $R$.
An example of such
topological theory is provided by the theory of Hurwitz numbers,
and the role of index $R$ can be played by different
structures, for instance, the Young diagrams in the Frobenius formula
(\ref{Frof}).
In this case, we
define the correlators
involving the sum over $R$:
\be\label{cor}
<i_1,\ldots,i_n > = <\phi_{i_1}(R)*\ldots *\phi_{i_n}(R)>=\sum_R d_R^2
\phi_{i_1}(R)\ldots \phi_{i_n}(R)
\ee
Now, it is clear that the equality
\be
\hat C_{ijk} \equiv
<<i,j,k>> = <\phi_i(R)*\phi_j(R)*\phi_k(R)*E[\beta,R]> = \nn \\
= \sum_m C_{ij}^m <\phi_m(R)*\phi_k(R)*E[\beta,R]> =
\sum_m C_{ij}^m \hat G_{mk}=\sum_m \hat C_{ij}^m G_{mk}
\ee
with
\be
E[\beta,R] = e_*^{\sum_i
\beta_i \phi_i(R)} \label{EBR}
\ee
continues to hold in this case.

As follows from the discussion above, the Hurwitz partition function
as a function of $\beta$
satisfies the {\it full} WDVV equations (\ref{WDVVf}). Sometime, for
restricted sets of $\beta$-variables, it happen to be also KP
$\tau$-functions \cite{I} but this is beyond our
consideration here. Instead we note that the weight $E[\beta,R]$
can be made more general than (\ref{EBR}), without changing anything
in the content of the previous consideration. Namely, one can change
(\ref{EBR}) for
\be
P_*(\phi(R)) *e_*^{\sum_i \beta_i\phi_i(R)}
\ee
with an arbitrary $*$-polynomial of observables $\phi_i(R)$ with the
same $R$. Sometimes new integrability properties can occur for the
partition function as a function on these additional parameters \cite{I}.

In Hurwitz theory {\it per se}, the role of observables $\phi_i(R)$
is played by the characters of the symmetric group $S_\infty$, denoted
by $\varphi_R(\Delta)$, where the label $i\rightarrow \Delta$ is now
the Young diagram. The most interesting choice for $P_*(\phi(R))$ is a
product of several $GL(\infty)$ characters,
$\chi_R(t)\chi_R(t')\chi_R(t'')\ldots$, where $\chi_R(t)$ is related to
$\varphi_R(\Delta)$ by formula (\ref{chiphitr}). With so
modified $E[\beta,R]$ the Hurwitz partition function becomes a
function of both $t$- and $\beta$-variables. While in
$\beta$-variables (when considering their complete set, not a subset)
it is usually a "quasiclassical $\tau$-function",
i.e. a solution to the full WDVV equations, in $t$-variables it can be a
KP $\tau$-function. This is, indeed, the case when there is one $(t)$
and two $(t,t')$ sets of $t$-variables, see \cite{GKM2,OkToda,I,III}.
Surprisingly or not, the KP integrability in
$t$ {\it dis}appears for three $(t,t',t'')$ or more sets of
$t$-variables. This peculiar pattern of
(non)-integrability structures is discussed in the next paper of this series \cite{III}.

In variance with our construction in this section,
in \cite{NTWDVV} there was suggested another, polynomial
class of solutions to the WDVV equations. In the Hurwitz theory context this
would correspond to a power series instead of polynomial solutions.

\section{Two multiplications of Young diagrams}

As we mentioned above, there are two natural multiplications on the Young diagrams: one,
$\circ$-multiplication given by the composition of permutations, and the other one,
*-multiplication induced by the algebra of cut-and-join operators (\ref{caj}).

\subsection{$\circ$-Multiplication of Young diagrams
from composition of permutations}

The $\circ$-multiplication is given on the Young
diagrams (integer partitions)
labeling elements of the group algebra of the symmetric group, i.e.
the sum of all permutations\footnote{
The composition of permutations is done in Maple by the command {\large{\it mulperms}}
of the package {\large{\it group}}.}
from the corresponding conjugation class:
$$
[211] = (12) + (13) + (14) + (23) + (24) + (34)
$$
etc.
The number of items is denoted by $||\Delta||$.

The naive $\circ$-multiplication  of Young diagrams,
induced by the (non-commutative but associative) composition
of permutations is commutative and associative.
Of course, $||\Delta_1\circ\Delta_2|| = ||\Delta_1||\cdot||\Delta_2||$.

Examples of the multiplication tables for different symmetric groups
can be found in Appendix II.

\noindent

$\bullet$
For any $k$ the $\circ$-multiplication by $[1^k]$ acts like unity:
\be
[1^k]\circ \Delta = \Delta
\ \ \ \ \ \ \forall \Delta: \ |\Delta|=k
\ee

$\bullet$
Multiplication by $[2,1^k]$ can be deduced from the
cut-and-join property.
Namely, if permutations are written in the cyclic notations, then
permutation  $(12) \in S_{k+2}$ acts as follows:
\be
(12) \circ (12)K = K, \nn \\
(12) \circ (12C)K = (1C)K, \nn \\
(12) \circ (21C)K = (2C)K, \nn \\
(12) \circ (1C)K = (12C)K, \nn \\
(12) \circ (C)K = (12)(C)K, \nn \\
(12) \circ (1C_1)(2C_2)K = (1C_22C_1)K
\label{12mult}
\ee
where $C$ denotes any set of elements,
and $K$ any set of non-intersecting cycles
(of course, it is assumed that $1,2\notin C,C_1,C_2,K$
and $C,C_1,C_2\notin K$).
If for a given level $k$ all the "powers" $||\Delta||$
are known, this is enough to construct all the entries
$[2,1^k]\circ\Delta$ in the $\circ$-multiplication table:
the coefficient for each line in (\ref{12mult})
is given by the ratio of items of the given type
at the l.h.s. of (\ref{12mult}) to the "power" at
the r.h.s., multiplied by $||2,1^k||$. These rules are illustrated in manifest
examples of Appendix I.

\subsection{*-multiplication of Young diagrams}

The *-multiplication of Young diagrams is associated with the product of
the differential cut-and-join operators (\ref{caj}): if
\be
\hat W[\Delta_1] \hat W[\Delta_2] = \sum_{\Delta} C_{\Delta_1\Delta_2}^\Delta
\hat W[\Delta],
\ee
then
\be
\Delta_1*\Delta_2 = \sum_{\Delta} C_{\Delta_1\Delta_2}^\Delta \Delta
\label{D*D}
\ee
and thus it is commutative and associative.
The sums are actually finite:
the size of the diagrams $\Delta$ is restricted to
\be
{\rm max}(|\Delta_1|,|\Delta_2|) \leq |\Delta|
\leq |\Delta_1|+|\Delta_2|
\ee
As it was already mentioned in the Introduction,
these commuting $\hat W$-operators have all the $GL(\infty)$
characters as common eigenfunctions, while $S_\infty$ characters
are the corresponding eigenvalues (\ref{WchiI}).
Representations of $GL(\infty)$ characters through the first
and second Weyl formulas are associated to representations
of the cut-and-join operators in time and matrix variables.
It follows from (\ref{WchiI}) that the symmetric group characters
form the same commutative associative algebra:
\be
\varphi_R(\Delta_1)\varphi_R(\Delta_2)
= \sum_{\Delta} C_{\Delta_1\Delta_2}^\Delta \varphi_R(\Delta)
\ee
with the same $R$-independent structure constants
$C_{\Delta_1\Delta_2}^\Delta$.

\subsection{Connection between the two multiplications\label{twop}}

The *-multiplication (\ref{D*D}) can be expressed through
the $\circ$-multiplication of Young diagrams.
It is a rather long recursive formula,
but actually a very constructive one:
$$
\Delta_1*\Delta_2 =
\sum_{n={\rm max}(|\Delta_1|,|\Delta_2|)}^{|\Delta_1|+|\Delta_2|}
\{\Delta_1,\Delta_2\}_n
$$
\be
\{\Delta_1,\Delta_2\}_n=\sum_{\Delta: |\Delta|=n}
C_{\Delta_1\Delta_2}^\Delta\Delta
= \rho_{n-|\Delta_1|}(\Delta_1)\circ \rho_{n-|\Delta_2|}(\Delta_2)
- \sum_{k={\rm max}(|\Delta_1|,|\Delta_2|)}^{n-1}
\rho_{n-k}\left(\{\Delta_1,\Delta_2\}_k\right)
\label{De*De}
\ee
and $\rho_k$ is a lift of the Young diagram to the size $|\Delta|+k$,
achieved by adding $k$ unit length rows with additional
numeric factor: if $\Delta$ already has $r$ rows of the length $1$,
then
\be
\rho_k([\Delta]) = \frac{(r+k)!}{r!k!}[\Delta,\underbrace{1,\ldots,1}_k],
\ \ \ \ \ \ \ \ \hbox{or}\ \ \ \ \ \ \ \
\rho_k([\tilde\Delta,1^r]) = \frac{(r+k)!}{r!k!}[\tilde\Delta,1^{r+k}]
\ee
where $[\Delta]\equiv[\tilde\Delta,1^r]$ and $\tilde\Delta$ does not contains
unit rows.
According to this definition, $\rho_0(\Delta) = \Delta$.
Note that
\be
\rho_k\left(\rho_l(\Delta)\right) = \frac{(k+l+r)!}{k!l!r!}\,
[\Delta,1^{k+l}] \ \ \ \neq \ \ \
\rho_{k+l}(\Delta) = \frac{(k+l+r)!}{(k+l)!r!}\,[\Delta,1^{k+l}]
\ee
The highest term in the product (\ref{De*De}) is
\be
\{\Delta_1,\Delta_2\}_{|\Delta_1|+|\Delta_2|}
= C_{\Delta_1\Delta_2}^{[\Delta_1,\Delta_2]} [\Delta_1,\Delta_2]
\ee
and for $\Delta_1 = [k^{m_k}]$, $\Delta_2 = [k^{n_k}]$,
$[\Delta_1,\Delta_2] = [k^{m_k+n_k}]$
the combinatorial coefficient is
\be
C_{\Delta_1\Delta_2}^{[\Delta_1,\Delta_2]}
= \prod_k \frac{(m_k+n_k)!}{m_k!n_k!}
\ee
This follows from the definition (\ref{caj}) of the $\hat W$ operator \cite{MMN}.

If $\Delta_1*\Delta_2 = \sum_{\Delta} C_{\Delta_1\Delta_2}^\Delta
\Delta$, formula (\ref{De*De}) can be rewritten as
\be
\rho_{n-|\Delta|_1}
(\Delta_1) \circ \rho_{n-|\Delta|_2} (\Delta_2) = \sum_m
\rho_{n-m}\Big(\{\Delta_1,\Delta_2\}_m\Big)
 = \sum_{\Delta}
C_{\Delta_1\Delta_2}^\Delta \rho_{n-|\Delta|} (\Delta)
\ee
Expressed in terms of the generating functions $J_\Delta(u) =
\sum_{m=0}^\infty u^{|\Delta|+m} \rho_m(\Delta)$ this multiplication
formula becomes
\be
\oint J_{\Delta_1}(u) \circ
J_{\Delta_2}\left(\frac{v}{u}\right) \frac{du}{u} = \sum_{\Delta}
C_{\Delta_1\Delta_2}^\Delta J_\Delta(v) = J_{\Delta_1*\Delta_2}(v)
\label{JJJ}
\ee
Note that the contour integral over $u$ at the l.h.s.
selects diagrams of the same weight, so that the operation $\circ$ is
well defined.

Examples of the $*$-multiplication tables can be found in section 6 and in
Appendix II, here we
consider only the case of product $[1]*[\Delta]$ which will be of use for our further
consideration.

\subsection{Example of level $(1,m)$}

For $\Delta$ of the size $|\Delta|=m$, which already has $r$ columns of height $1$,
one gets
$$
[1]*[\Delta] = \{1,\Delta\}_m + \{1,\Delta\}_{m+1},
$$
$$
\{1,\Delta\}_m = \rho_{m-1}[1]\circ \Delta = m[1^m]\circ\Delta = m\Delta,
$$
\be
\{1,\Delta\}_{m+1} = \rho_m[1]\circ \rho_1[\Delta] - \rho_1\left(\{1,\Delta\}_m\right)
= (m+1)[1^{m+1}]\circ (r+1)[\Delta,1] -   m(r+1)[\Delta,1] = (r+1)[\Delta,1],
\ee
$$
\boxed{
\[1]*\Delta = |\Delta|\,\Delta + (r+1)[\Delta,1]
}
$$
In other words, if $\Delta = [\tilde\Delta,1^r]$,
where $\tilde\Delta$ contains no more units, then
\be
[1] * [\tilde\Delta,1^r] = (|\tilde\Delta|+r)[\tilde\Delta,1^r]
+ (r+1)[\tilde\Delta,1^{r+1}]
\ee

\section{Generating functions}

Introduce now a linear form (average) on the Young diagrams:
\be
< \Delta > = \frac{\delta(\Delta,[1^{|\Delta|}])}{|\Delta|!} =
\sum_R d_R^2\varphi_R(\Delta) \delta_{|R|,|\Delta|}
\label{avedef}
\ee
It can be used to construct a variety of generating functions
for averages of Young diagrams products (correlators).

In fact, the projection to $|R|=|\Delta|$ in the sum over $R$
in (\ref{avedef}) can be eliminated, and
the {\it infinite} sum
\be
\sum_R d_R^2\varphi_R(\Delta)
= \sum_{R: \ |R|\geq|\Delta|} d_R^2\varphi_R(\Delta)
= e<\Delta>
\label{eave}
\ee
where $e = 2.718\ldots$
This formula is important for evaluation of the
partition functions in the case of *-products.

\subsection{The standard Hurwitz partition function}

The standard generating function of Hurwitz numbers is
\be
Z_\circ\{\beta_{\tilde\Delta}\}
= \sum_n q^n Z_\circ^{(n)} \{\beta_{\Delta}\} =
\sum_n q^n \left< \exp_\circ \left(\sum_{\tilde\Delta:\
|\tilde\Delta|\leq n} \beta_{\tilde\Delta}
\rho_{n-|\tilde\Delta|}(\tilde\Delta) \right) \right>\ =
\sum_n q^n \left< \exp_\circ \left(\sum_{\tilde\Delta:\
|\tilde\Delta|\leq n} \beta_{\tilde\Delta}
[\tilde\Delta,1^{n-|\Delta|}]\right) \right>
\label{circHPF}
\ee
where, as before, $\tilde\Delta$ denotes the Young diagram without units and
the $\circ$-multiplication of the Young diagram of different sizes is defined to
be zero. Actually, $q = e^{\beta_1}$.
Of course, one can also introduce a whole infinite
tower of $\beta$-variables for each $\tilde\Delta$:
$\beta_{\tilde\Delta,p} = \beta_{[\tilde\Delta,1^p]}$,
but we prefer not to do it.
Again, for each given $n$ the component
$Z_\circ^{(n)} \{\beta_{\Delta}\}$ satisfies the WDVV equations,
but after summation over $n$, (\ref{circHPF})
is not an average of any CAA exponential
and does not need to satisfy WDVV equations.
And, indeed, it does not.

With the help of (\ref{chiphitr}) one can rewrite (\ref{circHPF})
in terms of symmetric group characters $\varphi_R(\Delta)$, (\ref{phiext}).
Indeed, because of (\ref{WchiI}) they form a representation of the CAA algebra
with *-product:
\be
\varphi_R(\Delta_1)\varphi_R(\Delta_2)
= \varphi_R(\Delta_1*\Delta_2) =
\sum_{\stackrel{\Delta}{{\rm max}(|\Delta_1|,|\Delta_2|)
\leq |\Delta| \leq |\Delta_1|+|\Delta_2|}}
C^\Delta_{\Delta_1\Delta_2} \varphi_R(\Delta)
\ \ \ \ \ \ \ \ \forall \ R,\Delta_1,\Delta_2
\ee
For $|\Delta_1|=|\Delta_2|=n$ and for $|R|=n$ the property
(\ref{phiext}) implies that also
\be
\varphi_R(\Delta_1\circ\Delta_2) = \varphi_R(\Delta_1)\varphi_R(\Delta_2),
\ \ \ {\rm for}\ |R|=|\Delta_1|=|\Delta_2|
\label{circvarp}
\ee
and this allows one to express (\ref{circHPF}) through $\varphi_R(\Delta)$:
if all the sizes $|\Delta_1| = \ldots |\Delta_k| = n$ are the same, then
\be
\left< \circ_{i=1}^k \Delta_i \right> =
\sum_\Delta c^\Delta_{\Delta_1,\ldots,\Delta_k} <\Delta > \delta_{|\Delta|,n}
= \sum_\Delta c^\Delta_{\Delta_1,\ldots,\Delta_k} \sum_R d^2_R\varphi_R(\Delta)
\delta_{|\Delta|,n} \delta_{|R|,n}
= \sum_R d^2_R \prod_{i=1}^k \varphi_R(\Delta_i)\delta_{|R|,n}
\label{Wvsphi}
\ee
Here
\be
c^\Delta_{\Delta_1,\ldots,\Delta_k} = \sum_{\Delta'_1,\ldots,\Delta'_{k-2}}
C^\Delta_{\Delta_1\Delta'_1}
C^{\Delta'_1}_{\Delta_2\Delta_2'}\ldots C^{\Delta'_{k-2}}_{\Delta_{k-1}\Delta_k}
\delta_{|\Delta'_1|,n}\ldots\delta_{|\Delta'_{k-2}|,n}
\ee
we denote it by {\it small} letter $c$ to emphasize that all sums are restricted
(projected) to diagrams of the same size $n$.
Restriction to $|R|=n$ is important for the last transition in (\ref{Wvsphi}),
where {\it small} $c$ stand at the l.h.s.: still equality takes place
because of (\ref{circvarp}).

From (\ref{Wvsphi}) it follows directly that
partition function (\ref{circHPF}) can be rewritten as
\be
Z_\circ\{\beta_{\tilde\Delta}\} =
\sum_n q^n \sum_{R: \ |R|=n} d_R^2
\exp \left(\sum_{\tilde\Delta: \ |\tilde\Delta|\leq n}
\beta_{\tilde\Delta} \varphi_R(\tilde\Delta,1^{n-|\tilde\Delta|})\right)
\ee
or simply
\be
Z_\circ\{\beta_{\Delta}\} = \sum_n q^n Z_\circ^{(n)} \{\beta_{\Delta}\}
= \sum_n q^n \sum_{R: \ |R|=n} d_R^2
\exp \left(\sum_{\Delta: \ |\Delta| = n}
\beta_{\Delta} \varphi_R(\Delta)\right)
\ee
where, in principle, one can either impose the restriction
\be
\beta_{[\tilde\Delta,1^r]} = \beta_{\tilde\Delta}
\label{redbeta}
\ee
or not.

If (\ref{redbeta}) is not imposed, then,
making use of (\ref{chiphitr}),
one can further perform a Fourier transform of its $m$-th derivative
into $t$-variables:
\be
\sum_n q^n
\sum_{\stackrel{\Delta_1,\ldots,\Delta_m}{|\Delta_1|=\ldots=|\Delta_m|=n}}
\frac{\p^m Z_\circ^{(n)}\{\beta_\Delta\}}
{\p\beta_{\Delta_1}\ldots\p\beta_{\Delta_m}}
p^{(1)}(\Delta_1)\ldots p^{(m)}(\Delta_m) = \nn \\ =
\sum_n q^n
\sum_{R: \ |R|=n} d_R^{2-m}\chi_R(t^{(1)})\ldots\chi_R(t^{(m)})
\exp \left(\sum_{\tilde\Delta: \ |\tilde\Delta|=n}
\beta_{\tilde\Delta} \varphi_R(\tilde\Delta)\right)
\label{mchar}
\ee

\subsection{Extension to *-product}

With the *-product one can associate another, generalized
Hurwitz partition function \cite{MMN}:
\be
Z_*\{\beta_\Delta\} =
\left< \exp_*\left(\sum_\Delta \beta_\Delta \Delta\right)\right>
\ee
In variance with $Z_\circ\{\beta_\Delta\}$ in (\ref{circHPF}),
it satisfies the WDVV equations,
but in variance with individual components $Z_\circ^{(n)}\{\beta_\Delta\}$
(which also satisfy WDVV) it involves infinitely many time-variables
$\beta_\Delta$.

One can also rewrite $Z_*$ in terms of $\varphi_R(\Delta)$ characters,
but this time, in the case of *-products,
there would be no restriction on the sizes $|\Delta|$
in (\ref{Wvsphi}). Then, in this case the sum over $R$
will be restricted not to $|R|=n$, but to $|R|=|\Delta|$:
\be
\left< *_{i=1}^k \Delta_i \right> =
\sum_\Delta C^\Delta_{\Delta_1,\ldots,\Delta_k} <\Delta >
= \sum_\Delta C^\Delta_{\Delta_1,\ldots,\Delta_k} \sum_R d^2_R\varphi_R(\Delta)
\delta_{|R|,|\Delta|}
\label{Wvsphi*}
\ee
and, because of the $\Delta$-dependent projector,
in this formula one can not make any direct use of the relation
\be
\sum_\Delta C^\Delta_{\Delta_1,\ldots,\Delta_k}\varphi_R(\Delta)
= \prod_{i=1}^k \varphi_R(\Delta_i)
\label{Ccomb}
\ee
However, one can actually get rid of the projector!
The reason is that (\ref{chiphitr}) has important generalizations:
\be
\sum_\Delta d_R\varphi_R(\Delta)p\,(\Delta)\delta_{|\Delta|,|R|}
= \chi_R(t), \\
\sum_\Delta d_R\varphi_R(\Delta)p\,(\Delta)
= \chi_R(t_k+\delta_{k1}),
\label{chivarphiall} \\
\sum_{\tilde\Delta} d_R\varphi_R(\tilde\Delta)p\,(\tilde\Delta)
= \chi_R(1,t_2,t_3,\ldots),  \\
d_R = \chi_R(1,0,0,\ldots) = \chi_R(\delta_{k1})
\label{dRchar}
\ee
Note that  because of the property
(\ref{phiext}) all the sums over $\Delta$ are finite,
and these formulas are elementary, not transcendental.

Combining (\ref{chivarphiall}) and (\ref{dRchar}) with the celebrated
Cauchy completeness formula
\be\label{Cauchy}
\sum_R \chi_R(t)\chi_R(t') = \exp \left(\sum_k kt_kt'_k\right)
\ee
one obtains
\be
\sum_{R,\Delta} d_R^2 \varphi_R(\Delta) p(\Delta)
= \sum_R \chi_R(t_k+\delta_{k1})\chi_R(\delta_{k1}) =
e^{1+t_1}
= e \sum_{\Delta} <\Delta> p(\Delta)
= e \sum_{R,\Delta} d_R^2 \varphi_R(\Delta) p(\Delta)\delta_{|R|,|\Delta|}
\ee
In other words, we obtain the already-mentioned statement (\ref{eave}):
\be
<\Delta> = \sum_R d_R^2\varphi_R(\Delta) \delta_{|R|,|\Delta|}
= \frac{1}{e} \sum_R d_R^2\varphi_R(\Delta)
\ee
i.e. one can simply substitute everywhere the average (\ref{avedef}) by
the alternative one,
\be
<<\Delta>> = \sum_R d_R^2\varphi_R(\Delta) = e<\Delta>
\ee
where sum goes over Young diagrams $R$ of all sizes,
not restricted to $|R|=|\Delta|$.
The difference is actually exhausted by a factor of $e=2.718\ldots$

Coming back to (\ref{Wvsphi*}), one now knows how to eliminate
the unwanted projector from the r.h.s. and apply (\ref{Ccomb}):
\be
\left< *_{i=1}^k \Delta_i \right> = \frac{1}{e}
\left<\left< *_{i=1}^k \Delta_i \right>\right> = \frac{1}{e}
\sum_\Delta C^\Delta_{\Delta_1\ldots\Delta_k} <<\Delta >>
= \frac{1}{e}
\sum_\Delta C^\Delta_{\Delta_1\ldots\Delta_k} \sum_R d^2_R\varphi_R(\Delta)
= \frac{1}{e} \sum_R d^2_R\prod_{i=1}^k\varphi_R(\Delta_i)
\label{Wvsphi**}
\ee
In particular, the pair correlator is equal to
\be
\sum_{\Delta_1,\Delta_2}<\Delta_1*\Delta_2>p^{\Delta_1}\bar p^{\Delta_2}=
\sum_{\Delta_1,\Delta_2}\frac{1}{e} \sum_R d^2_R\varphi_R(\Delta_1)\varphi_R(\Delta_2)
p^{\Delta_1}\bar p^{\Delta_2}
\stackrel{(\ref{chivarphiall})}{=}\\=\frac{1}{e}
\sum_R\chi_R(t_k+\delta_{k1})\chi_R(\bar
t_k+\delta_{k1})\stackrel{(\ref{Cauchy})}{=}
\exp\left((t_1+1)(\bar t_1+1)-1+\sum_{k\ge 2}t_k\bar t_k
\right)\nn
\ee
From formula (\ref{Wvsphi**})
one obtains a character expansion of $Z_*$ and a much better
counterpart of (\ref{mchar}):
\be
Z_*\{\beta_\Delta\} =\
\left< \exp_*\left(\sum_\Delta \beta_\Delta \Delta\right)\right>\
= \frac{1}{e}\sum_R d^2_R \exp\left(\sum_\Delta \beta_\Delta \varphi_R(\Delta)\right)
\ee
and
\be
\sum_{\Delta_1,\ldots,\Delta_m}
\frac{\p^m Z_*\{\beta_\Delta\}}
{\p\beta_{\Delta_1}\ldots\p\beta_{\Delta_m}}
p^{(1)}(\Delta_1)\ldots p^{(m)}(\Delta_m) = \nn \\ =
{1\over e}\sum_n q^n
\sum_{R: \ |R|=n} d_R^{2-m}\chi_R(t^{(1)}_k+\delta_{k1})\ldots\chi_R(t^{(m)}_k+\delta_{k1})
\exp \left(\sum_{\Delta}
\beta_{\Delta} \varphi_R(\Delta)\right)
\label{mchar*}
\ee
We emphasize once again that there are no restrictions
on the sizes of any $\Delta_i$ and of $R$ in the sum.
In (\ref{mchar*}) we do not impose (\ref{redbeta}),
otherwise one should just write  correlators instead of
$\beta$-derivatives at the l.h.s.

\section{WDVV equations}

Whenever the generating function is an average
of the CAA exponential, it satisfies the WDVV equations.
This happens, at least, in two cases.
The first one is for the $\circ$-product at the given level $|\Delta|$:
then one gets a trivial WDVV solution in form of
a finite linear combination of ordinary exponentials.
The second case is that of the *-product: then the number of time-variables
is infinite, even in the simplest case of the $[1^p]$-subring,
an interesting open question being to find an adequate
quasiclassical (dispersionless) hierarchy, which is
associated with this particular solution to the WDVV equations:
it is probably related to the KP-Whitham hierarchy over
the Lambert curve \cite{mari,mmhk,kaz,Takasaki}.

We start with the trivial case of the $\circ$-partition function, then comment
on the case of the $*$-partition function when the WDVV equation are awaited on
common grounds but one tediously check them directly. We postpone until the next
two section a discussion of the case of $[1^p]$-subring when some
less involved direct checks
of the WDVV equations can be done.

\subsection{$\circ$-products at given level}

For a given $n$ one introduces a function of $|S_n|$
variables $\beta_\Delta$:
\be\label{wdvvtr}
Z_n(\beta) =  \left< \exp_\circ\left(
\sum_{\Delta:\ |\Delta|=n} \beta_\Delta \Delta\right) \right>
=
\sum_{R: \ |R|=n} d_R^2 \prod_{\Delta:\ |\Delta|=n}
e^{\beta_\Delta\varphi_R(\Delta)}
\ee
Each $Z_n$ satisfies the set of WDVV equations.
In this case this follows not only from the general arguments,
true for any topological field theory, but also
from a much simpler consideration:
one can obtain by a linear transform from $\{\beta_\Delta\}$ the separated variables
\be
\xi_R = \sum_{\Delta:\ |\Delta|=n} \beta_\Delta \varphi_R(\Delta)
\ee
so that
\be
\tilde Z_n(\xi) = \sum_R d_R^2e^{\xi_R} = Z_n(\beta)
\ee
Then the WDVV equations are trivially satisfied and, since this is no more than a linear
(and non-degenerate) change of beta-variables,
the original WDVV equations for $Z_n(\beta)$ are also true.

It is also evident that if one inserts into the sum (\ref{wdvvtr}) an arbitrary product of
characters $\chi_R(t^{(1)}_k)\ldots\chi_R(t^{(m)}_k)$, it does not spoil the
argument and, hence, the WDVV equations are still satisfied.

\subsection{The *-product case}

In this case one has to consider the WDVV equations with infinite matrices: there is
no any simple finite truncation for them. For instance, the simplest WDVV equation
$\check C_2\check C_3=\check C_3\check C_2$ (\ref{assC}) is
\be
\frac{<112><123>}{<11>} + \frac{<122><223>}{<22>} + \frac{<123><233>}{<33>} =
\frac{<113><122>}{<11>} + \frac{<123><222>}{<22>} + \frac{<133><223>}{<33>}
\ee
and it does {\it not} hold with
\be
<123> = <[1]*[2]*[3]> = 0, \ \ <113> = 0, \nn \\
<122> = 3/2, \ \ <223> = 1,  \ \ <133> = 4/3, \nn \\
<11> = 1, \ \ <22>=1/2, \ \ <33> = 1/3
\ee
Indeed, vanishing entries in the first line reduce the equation just to
\be
\frac{<122><223>}{<22>} = \frac{<133><223>}{<33>}
\ee
which is not true.

What is the reason? In fact, $\check C_2\check C_3=\check C_3\check C_2$ holds
in the following way:
\be
<(2*2)*3> = C^3_{22}<3*3> = C^3_{22}C_{33}^{111}<111> = 3\cdot 2\cdot\frac{1}{6}, \nn\\
<2*(2*3)> = C_{23}^{[21]}<2*[21]> = C_{32}^{[21]}C_{[21],2}^{111}<111>
= 3\cdot 2\cdot \frac{1}{6}
\ee
i.e. as  $\phantom._2(\check C_2\check C_3=\check C_3\check C_2)^{[111]}$:
\be
C^3_{22}C_{33}^{111} = C_{32}^{[21]}C_{[21],2}^{111}
\ee
However, while
\be
C^3_{22}C_{33}^{111} = \frac{<2*2*3>}{<3*3>}\frac{<3*3*111>}{<111*111>}
= \frac{<3*2*21>}{<21*21>}\frac{<2*21*111>}{<111*111>}= C_{32}^{[21]}C_{[21],2}^{111}
\ee
the symbolical expression neglecting the number of units does not hold:
\be
C_{22}^3C^1_{33}  \stackrel{?}{=} \frac{<223>}{<33>}\frac{<133>}{<11>}    \neq
\frac{<223>}{<22>}\frac{<122>}{<11>}  \stackrel{?}{=}  C^2_{32}C_{22}^1
\ee
Thus, reduction does not take place in the 1-sector.

Thus, one has to deal with infinite matrices. Though the WDVV equations still
have to be satisfied, checking them directly is a non-trivial problem. This is easier
to do in the simpler case of the $[1^p]$-subring case, which we discuss in the next
two sections.

\section{Sub-ring of $[1^p]$ operators and its action
on entire algebra}

In this section we discuss multiplication by the $[1^p]$ operators,
which form a closed sub-algebra of the entire algebra. Moreover, in
this case, it is possible to write down general formulas.

\subsection{*-subring of $[1^p]$ operators (single-line diagrams)
\label{1ring}}

The multiplication of $[1^p]$ operators is given by the formula
\be
[1^p]*[1^q]
= \sum_{i={\rm max}(0,p-q)}^p \frac{(q+i)!}{i!(p-i)!(q-p+i)!}\,[1^{q+i}]
= \sum_{s={\rm max}(p,q)}^{p+q} \frac{s!}{(s-p)!(s-q)!(p+q-s)!}\,[1^s]
\ee
Introduce $I(x) = \sum_p x^p[1^p]$. Then it follows that
\be
\boxed{
I(x) * I(y) = I(x+y+xy),} \nn \\
I(x)*I(y)*I(z) = I(x+y+xy)*I(z) = I(x+y+z+xy+yz+zx+xyz), \nn \\
\ldots, \nn \\
*_{i=1}^m I(x_i) =
I\left(\sum_i x_i + \sum_{i<j} x_ix_j + \sum_{i<j<k} x_ix_jx_k +\ldots
+ \prod_{i=1}^m x_i\right)
= I\left(-1 + \prod_i(x_i+1)\right)
\ee

\subsection{Action of $[1^p]$ operators on the entire algebra}

The multiplication of any operator by $[1^p]$ operators is described with the following
formulas.
Let $\Delta = [\tilde\Delta,1^r]$, with $1\notin \tilde\Delta$.
Let first $p\leq m = |\Delta|=|\tilde\Delta|+r$:
\be
[1^p]*[\Delta] = \sum_{i=0}^p\{1^p,\Delta\}_{m+i}, \nn \\
\{1^p,\Delta\}_m = \rho_{m-p}[1^p]\circ \Delta = \frac{m!}{p!(m-p)!}[1^m]\circ\Delta
= C^p_m\Delta, \nn \\
\{1^p,\Delta\}_{m+1} = \rho_{m-p+1}[1^p]\circ \rho_1[\Delta]
- \rho_1\left(\{1^p,\Delta\}_m\right) =\nn\\
= C^p_{m+1}[1^{m+1}]\circ (r+1)[\Delta,1]
-  C^p_{m}\cdot(r+1)[\Delta,1] = (r+1)C^{p-1}_m[\Delta,1], \nn \\
\{1^p,\Delta\}_{m+2} = \rho_{m-p+2}[1^p]\circ \rho_2[\Delta]
- \rho_2\left(\{1^p,\Delta\}_m\right) - \rho_1\left(\{1^p,\Delta\}_{m+1}\right)
=\nn\\
= C^p_{m+2}[1^{m+2}]\circ \frac{(r+1)(r+2)}{2}[\Delta,1,1]
- \frac{(r+1)(r+2)}{2}C^p_m [\Delta,1,1]
-  (r+2) \cdot(r+1)C^{p-1}_{m}[\Delta,1,1] = \nn \\
= \frac{(r+1)(r+2)}{2}\left(C^p_{m+2} - C^p_{m} - 2C^{p-1}_m\right)[\Delta,1,1]
= C^2_{r+2}C^{p-2}_m[\Delta,1,1], \nn \\
\ldots \nn \\
\boxed{
\[1^p]*\Delta = C^p_m\,\Delta + (r+1)C^{p-1}_m[\Delta,1] +
C^2_{r+2}C^{p-2}_m[\Delta,1,1] + \ldots
= \sum_{i=0}^p C_{r+i}^i C^{p-i}_m [\Delta,1^i]
}
\label{1p*}
\ee
In this form the formula holds also for $p>m$,
just the sum actually goes from $i={\rm max}(0,p-m)$.

Eq.(\ref{1p*}) can be rewritten also as
\be
\[1^p]*[\tilde\Delta,1^r] =
\sum_{i=0}^p C_{r+i}^i C^{p-i}_{|\tilde\Delta|+r} [\tilde\Delta,1^{r+i}]
\ee
and, further,
\be
I(x)*[\tilde\Delta,1^r] = \sum_{p,i} x^p C^{p-i}_{|\Delta|}C^i_{r+i}
[\tilde\Delta,1^{i+r}] =
(1+x)^{|\Delta|}\sum_i x^iC_{r+i}^i[\tilde\Delta,1^{i+r}]
\ee
If we introduce now a new generating function
$I_{\tilde\Delta}(x) = \sum_p x^p [\tilde\Delta,1^p]$, then
\be
\boxed{
I(x)*I_{\tilde\Delta}(y) = (1+x)^{|\tilde\Delta|}\sum_{i,r}
C_{r+i}^i x^i(1+x)^ry^r[1^{r+i}]
= (1+x)^{|\tilde\Delta|} I_{\tilde\Delta}(x+y+xy)
}
\ee
This formula describes the action of the $[1^p]$-subring
on the entire algebra of cut-and-join operators.

\subsection{$\circ$-correlators in the $[1^p]$ subring}

The $\circ$-multiplication is much simpler and in the
$[1^p]$ sector can be written by the two generating functions.
The first one is
\be
\sum_{n=0}^\infty q^n \left< [1^n]\circ [1^n]\right> t_1^n\bar t_1^n
=\sum_{n=0}^\infty q^n \left< [1^n]\right> t_1^n\bar t_1^n
=\sum_{n=0}^\infty \frac{q^n t_1^n\bar t_1^n}{n!} = \exp (qt_1\bar t_1)
\ee
Similarly,
\be
\sum_{n=0}^\infty q^n \left< [1^n]^{\circ m}\right> (t^{(1)}_1\ldots t^{(m)}_1)^n
= \exp \Big(qt^{(1)}_1\ldots t^{(m)}_1\Big)
\ee
The second generating function is
\be
\sum_{n=0}^\infty q^n \Big< \exp_\circ \left( \beta_{1^n}[1^n]\right)\Big>\
= \sum_{n=0}^\infty q^n \left< \delta_{n,0} + \beta_{1^n}[1^n] +
\frac{\beta_{1^n}^2}{2!}[1^n]\circ[1^n] + \ldots \right> =
\sum_{n=0}^\infty \frac{q^ne^{\beta_{1^n}}}{n!}
\label{beta1pcirc}
\ee
While each item in the sum is an average of a $\circ$-exponential,
the sum over $n$ is not, and one can {\it not} expect that {\it such}
partition functions satisfy the WDVV equations.

One can also consider a simplified version of
(\ref{beta1pcirc}), with all $\beta^{1^n}$ equal: $\beta_{1^n} = \beta_1$.
Then (\ref{beta1pcirc}) turns into
\be
\sum_{n=0}^\infty q^n \Big< \exp_\circ \left( \beta_{1}[1^n]\right)\Big>\
= \sum_{n=0}^\infty q^n \Big< \exp_\circ \left( \beta_{1}\rho_{n-1}[1]\right)\Big>\
= e^{\beta_1+q}
\ee
The standard Hurwitz partition function is direct generalization
of this formula.

\subsection{Connecting two multiplications}

Formula (\ref{JJJ}) connecting the two multiplications can be
further specified for the $[1^p]$ subring. Define in this case one
more generating function
\be
\sum_p x^p J_{[1^p]}(u) =
\sum_{p,m\geq 0} u^{p+m}x^p \rho_m([1^p]) = \sum_{m,p\geq 0}
\frac{(m+p)!}{m!p!} u^{p+m}x^p [1^{p+m}]
\ee
Then, in terms of $I(u) =
\sum_m u^m[1^m]$, one has: $J_{[1^p]}(u) = u^p\partial_u^p I(u)/p!$ and
one can easily relate (\ref{JJJ}) to $I(x)*I(y) = I(x+y +xy)$, so
that
\be
\oint J_{I(x)}(u)\circ J_{I(y)}\left(\frac{v}{u}\right)
\frac{du}{u} = J_{I(x+y+xy)}(v)
\ee
Note that in the generic case it would be
interesting to consider a generating function with the full set of
time variable $\{p_k\}$, $J(u|p) = \sum_\Delta J_\Delta(u)p_\Delta$,
so that
\be \oint J(u|p)\circ J\left(\frac{v}{u}\Big|\bar p\right)
\frac{du}{u} = \sum_{\Delta_1,\Delta_2} J_{\Delta_1*\Delta_2}(v)
\ee
and realize if there is an interesting expression for the r.h.s.

\section{Generalized Hurwitz partition function for
the $[1^p]$ subring}

\subsection{*-correlators}

Averaging converts $I(x)$ into the exponential:
\be
\Big< I(x) \Big> = \sum_p \frac{x^p}{p!} = e^x, \nn \\
\Big< *_i I(x_i) \Big> = \exp\left(-1 + \prod_i(x_i+1)\right)
\ee
or
\be
1+ \log \Big< *_i I(x_i) \Big> = \prod_i(1+x_i)
\ee
In particular,
\be
\sum_{k,l=0}^\infty \left< [1^k]*[1^l] \right> t_1^k \bar t_1^l
=\ <I(t_1+\bar t_1 + t_1\bar t_1) > \ = \exp (t_1+\bar t_1 + t_1\bar t_1)
\ee
Similarly
\be
\sum_{k_1,\ldots,k_m =0}^\infty \left< [1^{k_1}]*\ldots
*[1^{k_m}]\right> (t^{(1)}_1)^{k_1}\ldots (t^{(m)}_1)^{k_m} =
\exp\left(-1 + \prod_{i=1}^m \left((1+ t^{(i)}_m\right)\right)
\ee

\subsection{Correlators of *-exponentials}

It is convenient to introduce a grading of the diagram by its rescaling with a
formal parameter $q$: $[1^p]\to q^p[1^p]$.
The rescaled diagrams
(operators $q^p\hat W[1^p]$) also form a *-ring, but with rescaled
structure constants.
In terms of the generating function
$I_q(x) = \sum_p x^p q^p[1^p] = I(qx)$ one has
$I_q(x) * I_q(y) = I(qx)*I(qy) = I_q(x+y+qxy)$.
Note that for the $\circ$-multiplication $[1]^p\circ [1^p] = [1^p]$ and
logarithm of the average $\ \log\left<\sum_{p,q} x^py^q[1]^p\circ [1^q]
\delta_{p,q}\right> = xy$ is obtained from
$\ \displaystyle{{1\over q} \log <I_q(x)*I_q(y)>
={x+y\over q}+xy}\ $ in the limit of $q\rightarrow\infty$.

In order to check the associativity, one has to define
the partition function
\be
Z_*\{\beta|q\}= \left<
\exp_*\left(\sum_p \beta_{[1^p]}q^p[1^p]\right) \right> \ = 1 +
\sum_p \frac{1}{p!}q^p\beta_p + \frac{1}{2!}\sum_{p_1,p_2}
\beta_{p_1}\beta_{p_2}
\oint\frac{dx}{x^{p_1+1}}\oint\frac{dy}{y^{p_2+1}}e^{x+y+qxy} +
\ldots
\ee
and check if its third derivatives w.r.t. $\beta$'s satisfy the WDVV equations.

This quantity can be studied using the technique
developed above. For checks of the equation, one can use the perturbative
expansion of $Z_*\{\beta|q\}$ into the power series in $q$. Let us see how this works
in the leading order.

\subsection{WDVV equations for the $[1^p]$ subring}

Despite even for this subring the partition function depends on infinitely many
variables and, hence, the matrices of the third derivatives
\be\label{strc}
\left(\hat C_i\right)_{jk}
\equiv \hat C_{ijk}=C_{ijk} =
\frac{\p^3Z_*\{\beta|q\}}{\p\beta_i\p\beta_j\p\beta_k}
\ee
is infinite-dimensional, the associativity equations can be explicitly checked.

In the leading order approximation one has to check the associativity of the
non-deformed structure constants, i.e. (\ref{strc}) calculated at all $\beta_k=0$.
The generating functions for the structure constants $A_{pq}^s$ of the subring,
\be
[1^p]*[1^q] = \sum_{s={\rm max}(p,q)}^{p+q} A_{pq}^s [1^s]
\ee
are given by
\be
A^s(x,y) = \sum_{p,q} A_{pq}^s x^py^q = (x+y+xy)^s
\ee
or
\be
A(x,y;u) = \sum_{p,q,s} A_{pq}^s \frac{x^py^q}{u^{s+1}} = \frac{1}{u-(x+y+xy)}
\ee
The associativity is guaranteed by the symmetricity of
\be
\sum_{p,q,t} x^py^qz^t \sum_s A_{pq}^s A_{st}^r =
\oint A(x,y;u)A^r(u,z)du = (x+y+z+xy+yz+xz+xyz)^r
\label{ass}
\ee
w.r.t. $x\leftrightarrow z$ and $y\leftrightarrow z$.

Associativity condition (\ref{ass}) should be complemented by
\be
<[1^p]*[1^q]*[1^r]> = \sum_s A_{pq}^s <[1^s]*[1^r]>, \nn \\
e^{x+y+z+xy+yz+xz+xyz} = \oint A(x,y;u) e^{u+z+uz} du
\ee
which proves the WDVV equations in the leading order.

Further, one can switch on perturbations and check
the WDVV equations explicitly in higher orders. We certainly know that they
are correct basing on the general grounds, however, the procedure described here
allows one to check this iteratively.

\section*{Acknowledgements}

Our work is partly supported by Russian Federal Nuclear Energy
Agency under contract H.4e.45.90.11.1059, by Ministry of Education and Science of
the Russian Federation under contract 02.740.11.0608, by Russian government grant
11.G34.31.005, by RFBR
grants 10-02-00509 (A.Mir.), 10-02-00499 (A.Mor.) and 11-01-00289 (S.N.),
by joint grants 11-02-90453-Ukr, 09-02-93105-CNRSL, 09-02-91005-ANF,
10-02-92109-Yaf-a, 11-01-92612-Royal Society and by grant NSh 8462.2010.1.

\newpage

\part*{Appendices}

\section*{Appendix I. Multiplications of compositions.}

We illustrate the use of formulae of s.3 with the following example:
$\[2,1^r]\circ [2,1^r]$. Here $[2,1^r] = (12) + \ldots$
is a sum of $||2,1^r|| = C_{r+2}^2 = \frac{(r+1)(r+2)}{2}$
permutations. In the $\circ$-product each of them act on the
permutations in another $[2,1^r]$, and the action of, say,  $(12)$
by the rules (\ref{12mult}) is different on $(12)$, $(13)$ and $(34)$.
Then, on $(14)$ it is the same as on $(13)$ and on $(56)$ -- the
same as on $(34)$ -- the same in the sense that it produces
an element in the same conjugation class.
According to (\ref{12mult}),
\be
(12)\circ(12) = () \in [1^{r+2}], \nn \\
(12) \circ (13) = (123) \in [3,1^{r-1}], \nn \\
(12) \circ (34) = (12)(34) \in [22,1^{r-2}]
\ee
It remains to calculate the multiplicities:
\be
||2,1^{r-1}|| \cdot 1 /||1^{r+2}|| = ||2,1^r||
= C_{r+2}^2 = \frac{(r+1)(r+2)}{2},\nn\\
||2,1^{r-1}|| \cdot 2r /||3,1^{r-1}|| =
C_{r+2}^2\cdot 2r/2C_{r+2}^3 = 3, \nn \\
||2,1^{r-1}|| \cdot C_r^2 /||22,1^{r-2}|| =
C_{r+2}^2\cdot C_r^2/(C_{r+2}^2C_r^2/2) = 2
\ee
Here $1$, $2r$ and $C_r^2$ are the numbers of permutations
of the types $(12)$, $(1k)$ or $(2k)$ with $k\neq 1,2$ and
$(kl)$ with $k,l\neq 1,2$ in the conjugation class $[2,1^r]$.
Of course, $1 + 2r + C_r^2 = ||2,1^r|| = C_{r+2}^2$.
Thus we obtain:
\be
\boxed{
\[2,1^r]\circ [2,1^r] = \frac{(r+1)(r+2)}{2}[1^{r+2}]
+ 3[3,1^{r-1}] + 2[22,1^{r-2}]
}
\ee
Similarly,
\be
\boxed{
\[2,1^r]\circ [3,1^{r-1}] = 2r[2,1^r] + [32,1^{r-3}] + 4[4,1^{r-2}]
}
\ee
and so on.

$\bullet$
Similarly, for $[3,1^m]$:
\be
(123)\circ (123)K = (132)K, \nn \\
(123)\circ (132)K = K, \nn \\
(123)\circ (12C)K =(1C)(23)K, \nn \\
(123)\circ (13C)K=(12C)K, \nn \\
(123)\circ(1C)K = (123C)K, \nn\\
(123)\circ (C)K = (123)CK, \nn \\
(123)\circ (12C_1)(3C_2)K = (12C_13C_2) K\nn \\
(123)\circ (13C_1)(2C_2)K = (13C_12C_2) K  \nn \\
(123)\circ (1C_1)(2C_2)(3C_3)K = (1C_12C_23C_3) K
\label{123mult}
\ee
Therefore
\be
[3111]\circ[3111] = 40[1^6]+8[2211]+10[3111]+2[33]+5[51]
\ee
and in general
\be
[3,1^r]\circ [3,1^r] = \frac{2C_{r+3}^3}{2C_{r+3}^3}[3,1^r]
+ \frac{2C_{r+3}^3}{1}[1^{r+3}]
+ \frac{2C_{r+3}^3\cdot 3r}{C^2_{r+3}C^2_{r+1}/2}[22,1^{r-1}] + \nn \\
+ \frac{2C_{r+3}^3\cdot 3r}{2C_{r+3}^3} [3,1^r]
+ \frac{2C^3_{r+3}\cdot6C^2_{r}}{24C^5_{r+3}} [5,1^{r-2}]
+ \frac{2C^3_{r+3}\cdot 2C^3_{r}}{2C^3_{r+3}\cdot 2C^3_r/2}[33,1^{r-3}] = \nn \\
= \frac{(r+3)(r+2)(r+1)}{3}[1^{r+3}] + 8[22,1^{r-1}] +
(3r+1)[3,1^r] + 8[33,1^{r-3}] + 5[5,1^{r-2}]
\ee
A check of multiplicities in this formula:
\be
1+1+2\cdot 3r + 6C_{r}^2 + 2C^3_{r} = 2C^3_{r+3}
\ee

$\bullet$
The same formulas (\ref{123mult}) can be used to handle more general
$\circ$-products involving the triple cycles.
In particular,
\be
[3111]\circ [33] = 2[311] + 2[33] + 12[6], \nn \\
\[3,1^4]\circ[331] =  8[3,1^4] + 2[331] + 8[43] + 18[61]
\ee
and in general
\be
[3,1^{r+3}]\circ [33,1^r] = \frac{(r+1)(r+2)(r+3)}{3}\,[3,1^{r+3}]
+ 2[33,1^r] + 3[333,1^{r-3}] + \nn\\
+ 8[43,1^{r-1}] + 5[531^{r-2}] + 6(r+2)[61^r]
\ee

\section*{Appendix II. Multiplication tables}

\subsection{Group $S(2)$}

\be ||11|| = 1, \ \ \ ||2|| = 1 \ee

\be
\begin{array}{|c|c|}
\hline
[11] & [2] \\
\hline
[2] & [11] \\
\hline
\end{array}
\ee

\subsection{Group $S(3)$}

\be
\begin{array}{ll}
\[111] = (), & ||111|| = 1 \\
\[21] = (12) + (13) + (23),\ \ & ||21|| = 3 \\
\[3] = (123) + (132), & ||3|| = 2
\end{array}
\ee


\be
\begin{array}{|c|c|c|}
\hline
[111] & [21] & [3] \\
\hline
[21] & 3[111]+3[3] & 2[21] \\
\hline
[3] & 2[21] & 2[111]+[3] \\
\hline
\end{array}
\ee

\subsection{Group $S(4)$}

\be
\begin{array}{ll}
[1111] = (), & ||1111|| = 1 \nn \\
\[211] = (12) + (13) + (14) + (23) + (24) + (34), & ||211|| = 6 \nn \\
\[22] = (12)(34) + (13)(24) + (14)(23), &  ||22||=3 \nn \\
\[31] = (123)+(132) + (124)+(142) + (134) + (143) + (234) + (243),\ \  & ||31|| = 8 \nn \\
\[4] = (1234) + (1243) + (1324) + (1342) + (1423) + (1432), & ||4|| = 6
\end{array}
\ee


\be
\begin{array}{|c|c|c|c|c|}
\hline
[1111] & [211] & [22] & [31] & [4] \\
\hline
[211] & 6[1111]+2[22] + 3[31] & [211] + 2[4] & 4[211] + 4[4] & 4[22] + 3[31] \\
\hline
[22] & [211]+2[4] & 3[1111]+ 2[22] & 3[31] & 2[211] + [4] \\
\hline
[31] & 4[211] + 4[4] & 3[31] & 8 [1111] + 8[22] + 4[31] & 4[211] + 4[4] \\
\hline
[4]  & 4[22] + 3[31] & 2[211] + [4] & 4[211] + 4[4] & 6[1111] + 2[22] + 3[31] \\
\hline
\end{array}
\ee

\subsection{Group $S(5)$}

\be
\begin{array}{ll}
\[11111] = (), &  ||11111|| = 1 \\
\[2111] = (12) + \ldots, & ||2111|| = C^2_5 = 10 \\
\[221] = (12)(34) + \ldots, & ||221|| = \frac{1}{2}C^2_5 C^2_3 = 15 \\
\[311] = (123) + (132) + \ldots, & ||311|| = 2C^3_5 = 20 \\
\[32] = (123)(45) + \ldots, & ||32||=||311|| = 20 \\
\[41] = (1234) + \ldots, & ||41|| = 3!C^1_5 = 30 \\
\[5] = (12345) + \ldots  & ||5|| = 4! = 24
\end{array}
\ee

\be \sum_{\Delta:\ |\Delta|=5} ||\Delta|| = 5! = 120 \ee

\centerline{$
\begin{array}{|c|c|c|c|c|c|c|}
\hline
[11111] & [2111] & [221] & [311] & [32] & [41] & [5] \\
\hline [2111] & 10[11111]+2[221] + 3[311] & 3[2111]+3[32]+2[41]
&6[2111]+[32]+4[41]
& 4[221]+[311]+5[5] & 9[311]+5[5] & 8[41]\\
\hline
[221] & 3[2111]+3[32]+2[41] &&&&& \\
\hline
[311] &6[2111]+[32]+4[41] &&20[11111]+8[221]+&&& \\
&&&+7[311]+5[5] &&& \\
\hline
[32] &4[221]+[311]+5[5] &&&&& \\
\hline
[41]  &9[311]+5[5] &&&&&\\
\hline
[5] &8[41]&&&&& \\
\hline
\end{array}
$}

\section*{Appendix III. Structure constants of $*$-multiplication}

\subsection{Level $(1,1)$}

When it does not cause confusion we omit square brackets
in the notation of particular Young diagrams,
to simplify the formulas.
\be
[1]*[1] = \{1,1\}_1 + \{1,1\}_2, \nn \\
\{1,1\}_1 = [1]\circ[1] = [1], \nn \\
\{1,1\}_2 = \rho_1[1]\circ\rho_1[1] - \rho_1([1]\circ[1])
= \rho_1[1]\circ\rho_1[1] - \rho_1([1]) =
2[11]\circ 2[11] - 2\circ[11] = 2[11]
\ee
Thus
\be
\boxed{
[1]*[1] = \underline{[1]} + 2[11]
}
\ee
Here and below we underline the terms in the product,
which belong to the $\circ$-product:
they exist only when $|\Delta_1|=|\Delta_2|$
and are contained in the first term
$\{\Delta_1,\Delta_2\}_{|\Delta_1|=|\Delta_2|} = \Delta_1\circ\Delta_2$
with $n=|\Delta_1|=|\Delta_2|$
in the sum in (\ref{De*De}).

\subsection{Level $(1,2)$}

\be
[1]*[11] = \{1,11\}_2 + \{1,11\}_3, \nn \\
\{1,11\}_2 = \rho_1[1]\circ [11] = 2[11]\circ [11] = 2[11], \nn \\
\{1,11\}_3 = \rho_2[1]\circ \rho_1[11] - \rho_1\left(\{1,11\}_2\right)
= 3[111]\circ 3[111] - 2\cdot 3[111] = 3[111], \nn \\
\boxed{
\[1]*[11] = 2[11] + 3[111]
}
\ee
and
\be
[1]*[2] = \{1,2\}_2 + \{1,2\}_3, \nn \\
\{1,2\}_2 = \rho_1[1]\circ [2] = 2[11]\circ [2] = 2[2], \nn \\
\{1,2\}_3 = \rho_2[1]\circ \rho_1[2] - \rho_1\left(\{1,2\}_2\right)
= 3[111]\circ [21] - 2 [21] = [21], \nn \\
\boxed{
\[1]*[2] = 2[2] + [21]
}
\ee

\subsection{Level $(1,3)$}

\be
[1]*[111] = \{1,111\}_3 + \{1,111\}_4, \nn \\
\{1,111\}_3 = \rho_2[1]\circ [111] = 3[111]\circ[111] = 3[111], \nn \\
\{1,111\}_4 = \rho_3[1]\circ \rho_1[111] - \rho_1\left(\{1,111\}_3\right)
= 4[1111]\circ 4[1111] -  4\cdot 3[1111] = 4[1111], \nn \\
\boxed{
\[1]*[111] = 3[111] + 4[1111]
}
\ee

\be
[1]*[21] = \{1,21\}_3 + \{1,21\}_4, \nn \\
\{1,21\}_3 = \rho_2[1]\circ [21] = 3[111]\circ[21] = 3[21], \nn \\
\{1,21\}_4 = \rho_3[1]\circ \rho_1[21] - \rho_1\left(\{1,21\}_3\right)
= 4[1111]\circ 2[211] -  2\cdot 3[211] = 2[211], \nn \\
\boxed{
\[1]*[21] = 3[21] + 2[211]
}
\ee

\be
[1]*[3] = \{1,3\}_3 + \{1,3\}_4, \nn \\
\{1,3\}_3 = \rho_2[1]\circ [3] = 3[111]\circ[3] = 3[3], \nn \\
\{1,3\}_4 = \rho_3[1]\circ \rho_1[3] - \rho_1\left(\{1,3\}_3\right)
= 4[1111]\circ [31] -  3[31] = [31], \nn \\
\boxed{
\[1]*[3] = 3[3] + [31]
}
\ee

\subsection{Level $(2,2)$}

\be
[11]*[11] = \{11,11\}_2 + \{11,11\}_3 + \{11,11\}_4, \nn \\
\{11,11\}_2 = [11]\circ [11] = [11], \nn \\
\{11,11\}_3 = \rho_1[11]\circ \rho_1[11] - \rho_1\left(\{11,11\}_2\right)
= 3[111]\circ 3[111] -  3[111] = 6[111], \nn \\
\{11,11\}_4 = \rho_2[11]\circ\rho_2[11] - \rho_2\left(\{11,11\}_2\right)
- \rho_1\left(\{11,11\}_3\right) =
6[1111]\circ 6[1111] - 6[1111] - 4\cdot 6[1111] = 6[1111], \nn\\
\boxed{
\[11]*[11] = \underline{[11]} + 6[111]+6[1111]
}
\ee

\be
[11]*[2] = \{11,2\}_2 + \{11,2\}_3 + \{11,2\}_4, \nn \\
\{11,2\}_2 = [11]\circ [2] = [2], \nn \\
\{11,2\}_3 = \rho_1[11]\circ \rho_1[2] - \rho_1\left(\{11,2\}_2\right)
= 3[111]\circ [21] -  [21] = 2[21], \nn \\
\{11,2\}_4 = \rho_2[11]\circ\rho_2[2] - \rho_2\left(\{11,2\}_2\right)
- \rho_1\left(\{11,2\}_3\right) =
6[1111]\circ[211] - [211] - 2\cdot 2[211] = [211], \nn \\
\boxed{
\[11]*[2] = \underline{[2]} + 2[21]+[211]
}
\ee

\be
[2]*[2] = \{2,2\}_2 + \{2,2\}_3 + \{2,2\}_4, \nn \\
\{2,2\}_2 = [2]\circ [2] = [11], \nn \\
\{2,2\}_3 = \rho_1[2]\circ \rho_1[2] - \rho_1\left(\{2,2\}_2\right)
= [21]\circ [21] -  3[111] = 3[3], \nn \\
\{2,2\}_4 = \rho_2[2]\circ\rho_2[2] - \rho_2\left(\{2,2\}_2\right)
- \rho_1\left(\{2,2\}_3\right) =
[211]\circ[211] - 6[1111] - 3[31] = 2[22], \nn \\
\boxed{
\[2]*[2] = \underline{[11]} + 3[3] +2[22]
}
\ee

\subsection{Level $(1,4)$}

\be
[1]*[1111] = \{1,1111\}_4 + \{1,1111\}_5, \nn \\
\{1,1111\}_4 = \rho_3[1]\circ [1111] = 4[1111]\circ[1111] = 4[1111], \nn \\
\{1,1111\}_5 = \rho_4[1]\circ \rho_1[1111] - \rho_1\left(\{1,1111\}_4\right)
= 5[1111]\circ 5[1111] -  5\cdot 4[11111] = 5[11111], \nn \\
\boxed{
\[1]*[1111] = 4[1111] + 5[11111]
}
\ee

\be
[1]*[211] = \{1,211\}_4 + \{1,211\}_5, \nn \\
\{1,211\}_4 = \rho_3[1]\circ [211] = 4[1111]\circ[211] = 4[211], \nn \\
\{1,211\}_5 = \rho_4[1]\circ \rho_1[211] - \rho_1\left(\{1,211\}_4\right)
= 5[1111]\circ 3[2111] -  3\cdot 4[2111] = 3[2111], \nn \\
\boxed{
\[1]*[211] = 4[211] + 3[2111]
}
\ee

\be
[1]*[22] = \{1,22\}_4 + \{1,22\}_5, \nn \\
\{1,22\}_4 = \rho_3[1]\circ [22] = 4[1111]\circ[22] = 4[22], \nn \\
\{1,22\}_5 = \rho_4[1]\circ \rho_1[22] - \rho_1\left(\{1,2\}_4\right)
= 5[1111]\circ [221] -  4[221] = [221], \nn \\
\boxed{
\[1]*[22] = 4[22] + [221]
}
\ee

\be
[1]*[31] = \{1,31\}_4 + \{1,31\}_5, \nn \\
\{1,31\}_4 = \rho_3[1]\circ [31] = 4[1111]\circ[31] = 4[31], \nn \\
\{1,31\}_5 = \rho_4[1]\circ \rho_1[31] - \rho_1\left(\{1,31\}_4\right)
= 5[1111]\circ 2[311] -  2\cdot 4[311] = 2[311], \nn \\
\boxed{
\[1]*[31] = 4[31] + 2[311]
}
\ee

\be
[1]*[4] = \{1,4\}_4 + \{1,4\}_5, \nn \\
\{1,4\}_4 = \rho_3[1]\circ [4] = 4[1111]\circ[4] = 4[4], \nn \\
\{1,4\}_5 = \rho_4[1]\circ \rho_1[4] - \rho_1\left(\{1,4\}_4\right)
= 5[1111]\circ [41] -   4[41] = [41], \nn \\
\boxed{
\[1]*[4] = 4[4] + [41]
}
\ee

\subsection{Level $(2,3)$}

\be
[11]*[111] = \{11,111\}_3 + \{11,111\}_4 + \{11,111\}_5, \nn \\
\{11,111\}_3 = \rho_1[11]\circ [111] = 3[111], \nn \\
\{11,111\}_4 = \rho_2[11]\circ \rho_1[111] - \rho_1\left(\{11,111\}_3\right)
= 6[1111]\circ 4[1111] -  4\cdot 3[1111] = 12[1111], \nn \\
\{11,111\}_5 = \rho_3[11]\circ\rho_2[111] - \rho_2\left(\{11,111\}_3\right)
- \rho_1\left(\{11,111\}_4\right) = \nn \\ =
10[11111]\circ 10[11111] - 10\cdot 3[11111] - 5\cdot 12[11111] = 10[11111], \nn\\
\boxed{
\[11]*[111] = 3[111] + 12[1111]+10[11111]
}
\ee

\be
[11]*[21] = \{11,21\}_3 + \{11,21\}_4 + \{11,21\}_5, \nn \\
\{11,21\}_3 = \rho_1[11]\circ [21] = 3[21], \nn \\
\{11,21\}_4 = \rho_2[11]\circ \rho_1[21] - \rho_1\left(\{11,21\}_3\right)
= 6[1111]\circ 2[211] -  2\cdot 3[211] = 6[211], \nn \\
\{11,21\}_5 = \rho_3[11]\circ\rho_2[21] - \rho_2\left(\{11,21\}_3\right)
- \rho_1\left(\{11,21\}_4\right) = \nn \\ =
10[11111]\circ 3[2111] - 3\cdot 3[2111] - 3\cdot 6[2111] = 3[2111], \nn\\
\boxed{
\[11]*[21] = 3[21] + 6[211]+3[2111]
}
\ee

\be
[11]*[3] = \{11,3\}_3 + \{11,3\}_4 + \{11,3\}_5, \nn \\
\{11,3\}_3 = \rho_1[11]\circ [3] = 3[3], \nn \\
\{11,3\}_4 = \rho_2[11]\circ \rho_1[3] - \rho_1\left(\{11,3\}_3\right)
= 6[1111]\circ [31] -  3[31] = 3[31], \nn \\
\{11,3\}_5 = \rho_3[11]\circ\rho_2[3] - \rho_2\left(\{11,3\}_3\right)
- \rho_1\left(\{11,3\}_4\right) = \nn \\ =
10[11111]\circ [311] -  3[311] - 2\cdot 3[311] = [311], \nn\\
\boxed{
\[11]*[3] = 3[3] + 3[31]+[311]
}
\ee

\be
[2]*[111] = \{2,111\}_3 + \{2,111\}_4 + \{2,111\}_5, \nn \\
\{2,111\}_3 = \rho_1[2]\circ [111] = [21], \nn \\
\{2,111\}_4 = \rho_2[2]\circ \rho_1[111] - \rho_1\left(\{2,111\}_3\right)
= [211]\circ 4[1111] -  2[211] = 2[211], \nn \\
\{2,111\}_5 = \rho_3[2]\circ\rho_2[111] - \rho_2\left(\{2,111\}_3\right)
- \rho_1\left(\{2,111\}_4\right) = \nn \\ =
[2111]\circ 10[11111] -  3[2111] - 3\cdot 2[2111] = [2111], \nn\\
\boxed{
\[2]*[111] = [21] + 2[211]+[2111]
}
\ee

\be
[2]*[21] = \{2,21\}_3 + \{2,21\}_4 + \{2,21\}_5, \nn \\
\{2,21\}_3 = \rho_1[2]\circ [21] = [21]\circ[21] = 3[111]+3[3], \nn \\
\{2,21\}_4 = \rho_2[2]\circ \rho_1[21] - \rho_1\left(\{2,21\}_3\right)
= [211]\circ 2[211] -  4\cdot 3[1111] - 3[31] = 4[22]+3[31], \nn \\
\{2,21\}_5 = \rho_3[2]\circ\rho_2[21] - \rho_2\left(\{2,21\}_3\right)
- \rho_1\left(\{2,21\}_4\right) = \nn \\ =
[2111]\circ 3[2111] - 10\cdot 3[11111] - 3[311] - 4[221]-2\cdot 3 [311]
= 2[221], \nn\\
\boxed{
\[2]*[21] = 3[111] + 3[3] + 4[22] + 3[31] + 2[221]
}
\ee

\be
[2]*[3] = \{2,3\}_3 + \{2,3\}_4 + \{2,3\}_5, \nn \\
\{2,3\}_3 = \rho_1[2]\circ [3] = [21]\circ[3] = 2[21], \nn \\
\{2,3\}_4 = \rho_2[2]\circ \rho_1[3] - \rho_1\left(\{2,3\}_3\right)
= [211]\circ [31] - 2 \cdot 2[211] = 4[4], \nn \\
\{2,3\}_5 = \rho_3[2]\circ\rho_2[3] - \rho_2\left(\{2,3\}_3\right)
- \rho_1\left(\{2,3\}_4\right) = \nn \\ =
[2111]\circ [311] -  3\cdot 2[2111] - 4[41] = [32], \nn\\
\boxed{
\[2]*[3] = 2[21] + 4[4]+[32]
}
\ee

\bigskip

\bigskip

Above tables reproduce the ones from s.2.4.4 of \cite{MMN}.
What follows are some new pieces of the *-multiplication tables.

\subsection{Level $(1,5)$}

\be
\begin{array}{|lcl|}
\hline
\[1]*[5] &=& 5[5]+[51] \\
\[1]*[41] &=& 5[41]+2[411]  \\
\[1]*[32] &=& 5[32] + [321]  \\
\[1]*[311] &=& 5[311] + 3[3111]  \\
\[1]*[221] &=& 5[221]+2[2211]  \\
\[1]*[2111] &=& 5[2111] + 4[21111] \\
\[1]*[11111] &=& 5[11111] + 6 [111111] \\
\hline
\end{array}
\ee

\subsection{Level $(3,3)$}

\be
\boxed{
[111]*[111] = \underline{[111]} + 12[1111] + 30[11111] + 20[111111]
}
\ee

\be
\boxed{
[111]*[21] = \underline{[21]} + 6[211] + 9[2111] + 4[21111]
}
\ee

\be
\boxed{
[111]*[3] = \underline{[3]} + 3[31] + 3[311] + [3111]
}
\ee

\be
\boxed{
[21]*[21] = \underline{3[111]+3[3]} + 12[1111] + 8[22] + 9[31]
+ 10[221] + 6[311] + 4[2211]
}
\ee

\be
\boxed{
[21]*[3] = \underline{2[21]} + 4[211] + 8[4] + 3[32] + 4[41] + [321]
}
\ee

\be
[3]*[3] = \{3,3\}_3 + \{3,3\}_4 + \{3,3\}_5 + \{3,3\}_6, \nn \\
\{3,3\}_3 = [3] \circ [3] = 2[111]+[3], \nn \\
\{3,3\}_4 = \rho_1[3]\circ \rho_1[3] - \rho_1(\{3,3\}_3) =
[31]\circ[31] - 4\cdot 2[1111] - [31] =
 8[22] +3[31], \nn \\
\{3,3\}_5 = \rho_2[3]\circ \rho_2[3] - \rho_2(\{3,3\}_3) - \rho_1(\{3,3\}_4) =\nn\\
= [311]\circ [311] - 10\cdot 2[11111]-[311] - 8[221] - 2\cdot 3[311]
= 5[5], \nn \\
\{3,3\}_6 = \rho_3[3]\circ \rho_3[3] - \rho_3(\{3,3\}_3
- \rho_2(\{3,3\}_4) - \rho_1(\{3,3\}_5) = \nn \\ =
[3111]\circ [3111] - 20\cdot 2 [1^6] - [3111]
- 8[2211]-3\cdot 3[3111] - 5[51]
= 2[33], \nn \\
\boxed{
[3]*[3] = \underline{2[111]+[3]} + 8[22]+3[31] + 5[5]+2[33]
}
\ee

\subsection{Selected products}

\be
[3]*[33] = \{3,33\}_6 + \{3,33\}_7 + \{3,33\}_8 + \{3,33\}_9, \nn \\
\{3,33\}_6 = \rho_3[3]\circ [33] = [3111]\circ[33] = 2[3111]+2[33]+12[6],\nn\\
\{3,33\}_7 = \rho_4[3]\circ \rho_1[33] - \rho_1\left(\{3,33\}_6\right) = \nn \\
= [31^4]\circ[331] - 8[31^4] - 2[331] - 12[61] = 8[43] + 6[61], \nn \\
\{3,33\}_8 = \rho_5[3]\circ \rho_2[33] - \rho_2\left(\{3,33\}_6\right)
- \rho_1\left(\{3,33\}_7\right) = 5[53], \nn \\
\{3,33\}_9 = \rho_6[3]\circ \rho_3[33] - \rho_3\left(\{3,33\}_6\right)
- \rho_2\left(\{3,33\}_7\right) - \rho_1\left(\{3,33\}_8\right) = 3[333],\nn \\
\boxed{
[3]*[33]=2[33]+2[3111]+12[6]+8[43]+6[61]+5[53]+3[333]
}
\ee

\section*{Appendix III. The table of symmetric characters
$\varphi_R(\Delta)$ at level $6$}

\be
\begin{array}{|c|c||c|c|c|c|c|c|c|c|c|c|c|}
\hline
&&&&&&&&&&&&\\
R\ \ \Delta &d_R& [6] &[51]&[42]&[411]&[33]&[321]&[3111]&[222]&[2211]&[21111]&[111111]\\
&&&&&&&&&&&&\\
\hline\hline
&&&&&&&&&&&&\\
\lb 6\rb&\frac{1}{720}&120&144&90&90&40&120&40&{\bf 15}&45&15&1\\
&&&&&&&&&&&&\\
\hline
&&&&&&&&&&&&\\
\lb 51\rb&\frac{1}{144}&{\bf -24}&0&-18&18&-8&0&16&{\bf -3}&9&9&1\\
&&&&&&&&&&&&\\
\hline
&&&&&&&&&&&&\\
\lb 42\rb&\frac{1}{80}&0&-16&10&-10&0&0&0&{\bf 5}&5&5&1\\
&&&&&&&&&&&&\\
\hline
&&&&&&&&&&&&\\
\lb 411\rb&\frac{1}{72}&{\bf 12}&0&0&0&4&-12&4&{\bf -3}&-9&3&1\\
&&&&&&&&&&&&\\
\hline
&&&&&&&&&&&&\\
\lb 33\rb&\frac{1}{144}&0&0&-18&-18&16&24&-8&{\bf -9}&9&3&1\\
&&&&&&&&&&&&\\
\hline
&&&&&&&&&&&&\\
\lb 321\rb&\frac{1}{45}&0&9&0&0&-5&0&-5&{\bf 0}&0&0&1\\
&&&&&&&&&&&&\\
\hline
&&&&&&&&&&&&\\
\lb 3111\rb&\frac{1}{72}&-12&0&0&0&4&12&4&3&-9&-3&1\\
&&&&&&&&&&&&\\
\hline
&&&&&&&&&&&&\\
\lb 222\rb&\frac{1}{144}&0&0&-18&18&16&-24&-8&9&9&-3&1\\
&&&&&&&&&&&&\\
\hline
&&&&&&&&&&&&\\
\lb 2211\rb&\frac{1}{80}&0&-16&10&10&0&0&0&-5&5&-5&1\\
&&&&&&&&&&&&\\
\hline
&&&&&&&&&&&&\\
\lb 21111\rb&\frac{1}{144}&24&0&-18&-18&-8&0&16&3&9&-9&1\\
&&&&&&&&&&&&\\
\hline
&&&&&&&&&&&&\\
\lb 111111\rb&\frac{1}{720}&-120&144&90&-90&40&-120&40&-15&45&-15&1\\
&&&&&&&&&&&&\\
\hline
\end{array}
\ee

\end{document}